\documentstyle[eqsecnum,aps]{revtex}
\begin{document}
\draft
\title{Quantum Trajectory in Multi-Dimensional Non-Linear System}
\author{Hiroto Kubotani
\footnote{e-mail address: kubotani@yukawa.kyoto-u.ac.jp}}
\address{Faculty of Engineering, 
Kanagawa University, Yokohama 221-8686, Japan}
\date{\today}
\maketitle
\begin{abstract}
We discuss quantum dynamics in multi-dimensional non-linear systems.
It is well-known that wave functions are localized in a  
kicked rotor model.
However, coupling with other degrees of freedom breaks the 
localization. 
In order to clarify the difference, 
we describe  the quantum dynamics by deterministic rigid trajectories,  
which are accompanied with the de Broglie-Bohm interpretation of 
quantum mechanics, instead of wave functions.
A bundle of quantum trajectories are repulsive through quantum 
potential and flow never to go across each other.
We show that, according to the degrees of freedom, 
this same property appears differently.
\end{abstract}
\pacs{03.65.B2, 05.45.MT, 05.45.-a}
%
\section{Introduction}
%
When the question {\it \lq\lq What is quantum chaos?"} is thrown to us, 
we usually answer that no chaos is found in the quantum world. 
Successively, we cite a kicked rotor model \cite{Chirikov79} 
as a typical example.
Its classical dynamics is reduced to the standard map, 
which is a well-known chaotic system.
For a classical ensemble of the initial state, 
the dispersion of the momenta eternally 
continues to increase as a result of a random walk.
For a quantized kicked rotor, however, numerical simulation 
reveals that the diffusion in the momentum space
begins to be saturated after a time scale,  
and the wave function is found to be finally localized 
\cite{Chirikov81}\cite{Chirikov88}.
Non-existence of quantum chaos can 
be also explained in general terms.
For an isolated bound system, 
the energy spectrum is discrete and almost periodicity 
appears in the evolution of the wave function 
even if the energy spectrum is coarse-grained\cite{Percival61}. 
Further, even if the potential is allowed 
to be temporally periodic, the wave function of the bound states also 
evolves almost periodically as long as the system is not resonant\cite{Hogg82}.
In the proofs of these theorems, 
the property  that 
the Schr\"{o}dinger equation is a kind of wave equation 
and a wave function is a superposition of the oscillating solutions 
works as fundamental one to quantum mechanics.
Therefore, quantum chaos is occasionally called {\it quantum chaology} 
or {\it quantum psuedochaos}\cite{Chirikov88}\cite{Casati97}.
As reflection of this situation, 
the energy level statistics and the complex structure of eigenfunctions
in quantum non-linear systems 
\cite{Chirikov88} rather than the dynamics have been searched 
for remnants of classical chaos.

A decade ago, Adachi, Toda, and Ikeda\cite{Adachi88}\cite{Toda89}
analyzed a system of coupled kick rotors which 
are classically chaotic.
They found that localization of wave functions is broken in the model, 
while it appears when the coupling between the two rotors is absent.
Then, Ikeda, Adachi, and Toda\cite{Ikeda90} 
showed that a quantum system with helper degrees of freedom 
monotonously absorbs energy from the external field
\cite{Ikeda92}. 
Recently, Kubotani, Okamura, and Sakagami \cite{Kubotani95} 
have proposed that the coupled kicked rotors also yield
quantum noise with decaying correlation and break irreversibly the quantum 
coherence in contacted other degrees of freedom.
These results show that 
redundant degrees of freedom may drastically alter the quantum dynamics.
Further, if we are allowed to apply 
the Schr\"{o}dinger equation to the whole universe, 
which is isolated and composed of infinitely many degrees of 
freedom, 
the fundamental questions arises:
Why some degrees of freedom behave classically 
and others quantum-mechanically? 
Why only the formers are able to behave chaotically or irreversibly?
These problems cannot be answered straightforwardly \cite{Sakagami96}. 
Even without taking the ideal and mathematical limit of 
infinite degrees of freedom a priori,
therefore, the quantum dynamics still involves important 
and interesting phenomena, although 
chaos in the exact sense, which is believed to contribute to 
irreversibility in the classical world\cite{Zaslavsky84}, may not be 
present in the quantum world.

According to the de Broglie-Bohm interpretation of quantum mechanics
\cite{Holland93}, we can introduce a bundle of 
rigid trajectories which are consistently equivalent to a wave function.
At first sight, the rigidity may be inconsistent with the 
uncertainty principle in quantum mechanics.
Indeed, the uncertainty principle 
makes a product of the expected values of canonically 
conjugated variables superior to the Planck constant $\hbar$.
In the Copenhagen interpretation, the relation is considered 
as inaccuracy of the simultaneous measurement of the conjugate variables.
However, there is no mathematical inconsistency between the 
quantum trajectory picture and the time evolution of a wave function 
driven by the Schr\"{o}dinger equation 
as will be seen in Section 2.
The de Broglie-Bohm interpretation considers quantum expectation values 
as statistically averaged ones and thus  
the apparent inconsistency is due to the confusion of the 
measurement theories on which we base the interpretation of the 
uncertainty relation.

Although the recovery of classical diffusion is 
confirmed by the previous works \cite{Adachi88}-\cite{Sakagami96}, 
it does not necessarily mean that the classical dynamics 
is restored in quantum mechanics.
The dynamics needs to be estimated more directly.
The associated problems to be resolved are 
how nonlinearity work in quantum systems and 
what role redundant degrees of freedom play.
In order to reveal them, in this paper, 
we propose to utilize quantum trajectories.
For analysis of wave functions, indeed, 
the Wigner and Husimi functions have been used \cite{takahashi85}.
These projective representations 
of wave functions on the phase space can 
be compared with classical statistical distributions.
However, since they do not follow the Liouville equation, 
the comparison estimates only the correspondence or difference 
in the snap shot between
 the classical and quantum states.
For the analysis of the dynamics, therefore, the description of quantum 
mechanics by trajectories is favorable, 
since the trajectories can be easily compared with classical orbits.
By the quantum trajectory picture, even
the standard notions of chaos, such as the Lyapnov index and the KS entropy 
are also introduced  naturally into the quantum world
\cite{schwengelbeck95}.

This paper is organized as follows.
In the section 2, we introduce a coupled kicked rotator model as 
a multi-dimensional non-linear system and discuss the quantum picture of 
trajectories.
We calculate the quantum trajectory of the system in the section 3.
In the section 4, we give the summary and discuss the results.

%
\section{Multi-dimensional Non-Linear Model}
%
The kicked rotor model was first introduced by
Chirikov\cite{Chirikov79}.
Its property has been widely investigated 
both in classical and quantum mechanics, 
since it has the dynamics which is  typical of nonlinear systems.
The Hamiltonian is 
\begin{equation}
H_k(q,p;t)
=\frac{1}{2}p^2+k{\rm cos}(q)\sum_{n=1}^\infty\delta(t-n{\rm T})
~~~~(0 \le q < 2\pi),
\end{equation}
where $q$ and $p$ are an angle variable with period 2$\pi$ 
and its canonically conjugated action variable, respectively, and 
$k$ and $T$ are constant.
The first term is quadratic with respect to the action variable $p$ and 
shows a nonharmonic oscillator.
Alternatively we can also 
consider it as the kinetic term with respect to the momentum $p$.
The second one shows the temporally periodic and instantaneous kick, 
whose strength and period are parameterized by $k$ and $T$, respectively.
The classical dynamics is reduced to a discretized map.
The variables $(q_n,p_n)\equiv (q(t=nT+0),p(t=nT+0))$ 
 at the discretized time $t=nT+0$ ($n=0, 1, 2, \cdots$)
are transformed as
\begin{eqnarray}
&q_{n+1}=&q_n+T p_n, \nonumber \\
&p_{n+1}=&p_n+k{\rm sin}(q_{n+1}).
\label{standardmap}
\end{eqnarray}
Here $+0$ indicates the time just after the kick.
The transformation rule (\ref{standardmap}) is called the standard map, 
whose dynamics is characterized by one parameter $K\equiv kT$.
When $K \lesssim 1$, the variables $(q_n,p_n)$ are trapped on the KAM torus.
When $K$ exceeds the critical value $K_{cr}=0.9716\cdots$, they 
diffuse into the chaotic sea on the phase space.

We couple two kicked rotors whose kick strengths are $k_1$ and $k_2$.
The total Hamiltonian of the coupled rotors is given by
\begin{equation}
H(q_1,p_1,q_2,p_2;t)=H_{k_1}(q_1,p_1;t)+H_{k_2}(q_2,p_2;t)
+H_{int}(q_1,p_1,q_2,p_2;t).
\label{Hamiltonian}
\end{equation}
This type of model has been used 
by Adachi et al \cite{Adachi88}, Kubotani et al 
\cite{Kubotani95}, and Sakagami et al\cite{Sakagami96}.
In this paper we choose the interaction Hamiltonian 
$H_{int}$ as 
\begin{equation}
H_{int}(q_1,p_1,q_2,p_2;t)
=c_{pp}p_1 p_2, 
\label{Hint}
\end{equation}
where $c_{pp}$ 
is a coupling 
constant.
The interaction Hamiltonian $H_{int}$, Eq. (\ref{Hint}) 
causes continuous interaction between the two degrees of freedom 
through the action variables $p_1$ and $p_2$.
Hereafter we call the coupled kick rotors as a pp-coupling model.

We quantize the coupled kicked rotors (\ref{Hamiltonian}).
In the Shr\"{o}dinger picture, 
we write down the time evolution operator as  
\begin{equation}
\hat U(t)={\rm T_O}{\rm exp}[ 
                 -{i \over \hbar}\int_{0}^{t} ds \hat H(s)],
\label{Unitary}
\end{equation}
where ${\rm T_O}$ denotes the time ordered product, and 
$\hat H(t)$ is the quantized Hamiltonian operator 
which is derived by replacing the canonical variables $q_1$, $q_2$, $p_1$, 
and $p_2$ in Eq.(\ref{Hamiltonian}) with the corresponding 
quantum operators $\hat{q_1}$, $\hat{q_2}$, $\hat{p_1}$, and $\hat{p_2}$.
In general, incommutability between the conjugate operators 
in the Hamiltonian complicates estimation of the time ordered product.
For the pp-coupling model, 
we only have to take account of the ordering at 
the time when the kick is added.
As a result, the operator (\ref{Unitary}) is reduced to a simple form
\begin{equation}
\hat U(t)=\hat U_1(t_0)(\hat U_2 \hat U_1(T))^n,
\label{unitaryop}
\end{equation}
where
\begin{eqnarray}
\hat U_1(t)&\equiv&
{\rm exp}[-{i \over 2\hbar}(\hat p_1^2+\hat p_2^2
            +2c_{pp} \hat p_1 \hat p_2)t],\\
\hat U_2~~~&\equiv&
{\rm exp}[-{i \over \hbar}
(k_1 {\rm cos}(\hat q_1)+k_2 {\rm cos}(\hat q_2)
)].
\end{eqnarray}
Here $t_0\equiv t-n_0 T$ and $n_0=[t/T]$, where $[x]$ denotes the maximum 
integer which is not superior to $x$.
Note that the operators $\hat U_1$ and $\hat U_2$ 
are constructed only from the the action variable operators 
$\hat p_1$ and $\hat p_2$ 
or the angle variable operators $\hat q_1$ and $\hat q_2$, respectively, 
and can be expressed as diagonal matrices by choosing 
an appropriate representation.
Therefore, the operation of Eq. (\ref{unitaryop}) to a wave function is 
decomposed into the two procedures: 
the interchange of the representation of the wave function 
and the multiplication of complex factors.
The simplicity guarantees low roundoff error in 
the numerical simulation of a wave function.

According to the de Broglie-Bohm interpretation of quantum theory 
\cite{Holland93},
we construct a bundle of deterministic rigid trajectories from 
a wave function in the pp-coupling model.
First, we express the wave function with respect to the polar coordinate:
\begin{equation}
\Phi(q_1,q_2;t)=R(q_1,q_2;t) {\rm exp}[{{\rm i}\over \hbar}S(q_1,q_2;t)],
\end{equation}
where the $q_1,q_2$ representation is chosen, and 
$R$ and $S$ are real-valued functions.
Using this expression, we decompose the Schr\"{o}dinger 
equation into two parts:
\begin{equation}
{\partial \over \partial t}R^2 
+{\partial \over \partial q_1}
    R^2(({\partial S \over \partial q_1})
         +c_{pp}({\partial S \over \partial q_2}))
+{\partial \over \partial q_2}
    R^2(({\partial S \over \partial q_2})
         +c_{pp}({\partial S \over \partial q_1}))=0,
\label{continuity}
\end{equation}
and
\begin{equation}
{\partial S \over \partial t}
+{1 \over 2}({\partial S \over \partial q_1})^2
+{1 \over 2}({\partial S \over \partial q_2})^2
+c_{pp}({\partial S \over \partial q_1})({\partial S \over \partial q_2})
+V(q_1,q_2;t)+V_Q(q_1,q_2;t)=0,
\label{HamiltonJacobi}
\end{equation}
where $V(q_1,q_2;t)$ is a original potential:
\begin{equation}
V(q_1,q_2;t)
=(k_1{\rm cos}(q_1)+k_2{\rm cos}(q_2))
\sum_{n=1}^\infty\delta(t-n{\rm T}),
\end{equation}
and $V_Q(q_1,q_2)$ is what we call a quantum potential:
\begin{equation}
V_Q(q_1,q_2;t)
=-{1 \over 2R}({\partial^2 R \over \partial^2 q_1}
+{\partial^2 R \over \partial^2 q_2}
+2c_{pp}{\partial^2 R \over \partial q_1 \partial q_2})
\hbar^2.
\label{VQ}
\end{equation}

Next, we consider an ensemble 
consisting of quantum particles with the initial distribution function 
$R(q_1,q_2;t=0)^2$.
For the particles, we can define rigid trajectories as follows.
The momentum $(p_1, p_2)$ of a particle on a trajectory is equalized to 
the spatial gradient of the phase part of the wave function, $S$:
\begin{equation}
p_i=({\partial S \over \partial q_i})~~~~(i=1, 2).
\label{pdef}
\end{equation}
From the Hamilton's canonical equation, 
the particle moves according to the velocity:
\begin{eqnarray}
&\dot q_1=&p_1 + c_{pp}p_2
=({\partial S \over \partial q_1})+c_{pp}({\partial S \over \partial q_2}),
\nonumber \\
&\dot q_2=&p_2 + c_{pp}p_1
=({\partial S \over \partial q_2})+c_{pp}({\partial S \over \partial q_1}).
\label{velocity}
\end{eqnarray}
Integration of Eq. (\ref{velocity}) with respect to $t$ yields 
a time-parameterized trajectory.
The relation (\ref{velocity}) allows Eq. (\ref{continuity}) to be 
rewritten as
\begin{equation}
{\partial \over \partial t}R^2 
+{\partial \over \partial q_1}
   \dot q_1 R^2
+{\partial \over \partial q_2}
   \dot q_2 R^2  
        =0,
\label{rewrite}
\end{equation}
which reserves conservation of the distribution function $R(q_1,q_2;t)^2$.
We also note that Eq. (\ref{HamiltonJacobi}) corresponds to the 
Hamilton-Jacobi equation in classical mechanics.
Compared with the classical one, 
the additional term $V_Q$ appears in Eq. (\ref{HamiltonJacobi}).
The term shows quantum effect on quantum trajectories.
That is why $V_Q$ is called a quantum potential.
We mention that Eq. (\ref{pdef}) appears to show WKB approximation.
By definition, however, $S$ is not equivalent to the phase part of 
the WKB wave function, $S_{{\rm WKB}}$.
$S_{{\rm WKB}}$ is determined by the classical Hamilton-Jacobi equation 
which does not include the quantum potential $V_Q$.
Therefore, only if $V_Q$ is negligible,  $S$ happens to 
be equal to $S_{{\rm WKB}}$.
In other words, the quantum trajectory picture
is not limited to WKB regions and always defined 
consistently irrespective of any approximation schemes.

%
\section{Quantum Trajectory in Non-Linear System}
%
In this section, 
we numerically trace quantum trajectories in the coupled kick rotor model
constructed in the section 2 and compare them with classical ones.
For this aim, 
we first solve the original Schr\"{o}dinger equation instead of 
the quantum Hamilton-Jacobi equation (\ref{HamiltonJacobi}).
To represent a wave function numerically, 
we use $4096 \times 512$ mesh points.
Setting $\hbar=2\pi \times 43 / 4096=0.0659\cdots$, 
we can describe the momentum eigenstates with the eigenvalues $p_1$ 
from $-43\pi$ to $43\pi$ and $p_2$ from $-43\pi/8$ to $43\pi/8$.
For the model to analyze, we fix the parameters as
$k_1=2.0$, $k_2=0.9$, and ${\rm T}=1.0$.
As an initial wave function, 
we choose a product of the momentum eigenstates for 
each degree of freedom:
\begin{eqnarray}
\Phi_{0}(q_1,q_2)&=&\Phi_{{p_1}_0}(q_1)\otimes \Phi_{{p_2}_0}(q_2)
\nonumber \\
&=&{1 \over \sqrt{2\pi}}{\rm exp}(-{i \over \hbar}{p_1}_0
(q_1-{\pi \over 4}))
\times 
 {1 \over \sqrt{2\pi}}{\rm exp}(-{i \over \hbar}{p_2}_0
(q_2-{\pi \over 4})),
\label{Initial}
\end{eqnarray}
where ${p_1}_0$ and ${p_2}_0$ are initial eigenvalues.
In the following concrete calculation, we set ${p_1}_0={p_2}_0=\pi/2$.
The wave function at the arbitrary time is obtained by multiplying 
the initial wave function by the operator (\ref{unitaryop}).
From the time-dependent wave function, next, 
we estimate the velocity of a quantum particle, (\ref{velocity}).
The spatial derivative of the wave function just on the mesh point is 
estimated by the inverse Fourier transformation.
The value between the mesh points is interpolated\cite{recipes}.
A quantum trajectory is calculated by integrating 
the velocity field by the Rungge-Kutta method.

We begin by observing gross time evolution of quantum state 
with respect to the dispersion of the momentum $\hat p_1$.
Fig. 1 shows the second order moment $Q\equiv<(\hat p_1-<\hat p_1>)^2>$, 
where $<>$ denotes the expectation value estimated with respect to 
the wave function at the time $t=n{\rm T}+0$.
The lines (No) and (P-P) indicate the single kick rotor model 
($k_1=2.0$, $c_{pp}=0$) and the pp-coupling model ($k_1=2.0$, $k_2=0.9$, 
and $c_{pp}=0.2$), respectively.
For the single rotor, localization of a wave function is found. 
For the coupled kicked rotors, on the other hand, 
the diffusion in the momentum space seems to continue.
In Fig. 2, the distribution of the momentum $\hat p_1$ at $t=100{\rm T}+0$, 
$|<p_1|\Phi(t=100{\rm T}+0)>|^2$ is shown for each model.
The distribution (No) is  approximate to 
${\rm ln}|<p_1|\Phi>|^2 \propto -|p_1|$, 
although the classical law for the chaotic rotor would be 
${\rm ln}f(p) \propto -p^2$, where $f(p)$ is a classical distribution 
function.
Comparison with the quantum kicked rotor modulated by 
classical noise is discussed in \cite{Toda89}.

Fig. 1 shows that introduction of coupling with other degrees of freedom 
may change the quantum dynamics drastically.
However, the recovery of classical diffusion found here does not 
necessarily mean that the  Liouville dynamics is restored in 
the evolution of the wave function.
Fig. 2 shows the plateau with $2\pi$ period appears 
in the distribution for the coupled rotors.
As well as Fig. 1, 
it suggests similarity between the classical and quantum 
dynamics\cite{Toda89}.

The classical dynamics of a single kicked rotor is reduced to 
a map between the variables at discretized time 
as mentioned in the section 2.
For the coupled kicked rotors (\ref{Hamiltonian}), 
the map is 
\begin{eqnarray}
q_{i,n+1}=&q_{i,n}+&{\rm T}p_{i,n}+c_{pp}p_{j,n} ~~~~(i=1, 2, j\neq i),
\nonumber \\
p_{i,n+1}=&p_{i,n}+&k_i{\rm sin}(q_{i,n+1}), 
\label{2Dmap}
\end{eqnarray}
where $(q_{i,n},p_{i,n})=(q_i(t=n{\rm T}+0), p_i(t=n{\rm T}+0))$ 
are the canonical variables at the discretized time $t=n{\rm T}+0$ 
(n=0, 1, 2, $\cdots$). 
The classical particle in the region [0, $\pi$] is accelerated and 
one in the region [$\pi$,2$\pi$] is decelerated by a kick.
Until the next kick, the momentum is conserved and the velocity of the 
particle is constant.
The dynamics results in effective stretching and folding
of a phase space volume element.

For the quantum particle, on the other hand, 
the velocity can be altered, 
even if the original potential in the Hamiltonian is not present.
The instantaneous kick yields the inhomogeneous velocity field of the 
quantum particle.
After the kick, as a result, the particles are gathered inhomogeneously 
and high density regions are produced.
The gradient of the density of the particles gives nonzero quantum potential 
through Eq. (\ref{VQ}).
In other words, the quantum particle feels repulsive force 
from the other particles.
Further, uniqueness of the solution of the differential 
equation (\ref{velocity}) guarantees that two quantum trajectories
cannot go across each other.
Fig. 3 shows quantum trajectories just after the 1st kick.
The lines in Fig. 3(a) are trajectories for the single rotor model
 with no coupling.
At $t={\rm T}$, 20 particles are settled homogeneously between 
0 and $2\pi$ on the $q_1$-axis.
Figs. 3(b) and 3(c) show typical quantum trajectories 
for the pp-coupling model.
For Figs. 3(b) and 3(c), we set 20 particles evenly on 
the surfaces $q_2=\pi$ and $q_2=0$, respectively, at $t={\rm T}$.
The strongly repulsive nature of the trajectory 
is observed in Fig. 3(a).
Although the repulsion is also found in Fig. 3(b), for Fig. 3(c) it 
is partially milder than ones found in Figs. 3(a) and 3(b).  
For the coupled rotors, 
we also see the evolution of the quantum trajectory in the 2-dimensional 
configuration space $(q_1,q_2)$ after the 1st kick.
Fig. 4 shows the evolution of the 500 particles which are 
equally spaced on the surfaces $q_2 =\pi$ and 
$q_2 =0$ at $t={\rm T}$ as well as Figs. 3(b) and 3(c).
The initial straight row of particles is going 
to be broken at the region where 
the particle density is increased by condensation of the quantum 
trajectories.

Next, we estimate effective acceleration of a quantum particle 
induced by a kick.
In the classical system, nonlinearity in the kick is enhanced by  
the free motion after the kick.
For a kicked rotor, therefore, the efficiency of the nonlinearity 
is parameterized by $K \equiv k{\rm T}$, rather than bare kick strength $k$. 
In the quantum system, however, 
the free motion is suppressed by the quantum potential, 
as seen in Fig. 3.
To quantify the effect, 
we propose to estimate the averaged velocity $V_i(n)$ by 
the distance over which the quantum particle moves during the time  
interval $[nT, (n+1)T]$:
\begin{equation}
V_i(n)\equiv{1 \over T}(q_{i,n+1}-q_{i,n}).
\end{equation}
Further, the change rate of the effective velocity $V_i$ is 
considered as the effective strength of the impulsive kick.
Therefore, we estimate the effective kick strength by
\begin{equation}
F_i(n)\equiv{1 \over T}(V_i(n)-V_{i}(n-1))
={1 \over T^2}((q_{i,n+1}-q_{i,n})-(q_{i,n}-q_{i,n-1})).
\end{equation}
For the classical standard map (\ref{2Dmap}), $F_i$ is approximately 
 a sine function with respect to $q_{i,n}$.
For the quantum system, we calculate $F_1(n)$ numerically.
At the time $t=30{\rm T}$,
when the difference in the dispersion 
$<(\hat p_1-<\hat p_1>)^2>$ between the single rotor and the coupled ones 
is apparently observed(Fig.1), we let $500 \times 20$ probe particles 
distributed uniformly on the $q_1$-coordinate for the single kick 
rotor, and on the configuration space $(q_1, q_2)$ for the 
coupled kick rotor model.
Figs. 5(a) and 5(b) are the results for the single kicked rotor and 
 pp-coupling models, respectively.
For some particles, 
the numerical estimation of $V_1(29)$ and $V_1(30)$
does not converge with respect to the integration step size, 
since they may go through 
the irregularly large quantum potential.
In Fig. 5(b), we plot $F_1(30)$ for the $9899$ particles 
whose deviations in $V_1(29)$ and $V_1(30)$ are smaller than $0.1$ 
if we make the integration step size half.
For the coupled kick rotors, we can find the tendency that 
the particles in the region $0 <q_1 <\pi$ are accelerated and 
ones in the region $\pi <q_1 <2\pi$ are decelerated, 
although not a few particles don't follow the tendency and 
the acceleration and deceleration amplitude is weaker than 
in the genuine classical case $2.0~{\rm sin}(q_1)$.
The tendency is obscure in Fig.5(a) for a single rotor model.

We see the evolution of volume elements in the 2-dimensional 
configuration  space $(q_1,q_2)$.
At $t={\rm T}$, we distribute initial probe particles uniformly between 
$0$ and $2\pi$ on the surfaces $q_2=2\pi \times {i \over 10}$
(i=0, 1, $\cdots$, 9) as shown in Fig. 6(a).
Figs. 6(b) and 6(c) show snapshots on the configuration space ($q_1,q_2$) 
at $t=2{\rm T}$ and $t=3{\rm T}$, respectively.
The density of quantum particles is conserved in accordance with 
Eq. (\ref{continuity}).
The volume elements of the density in the configuration space 
obtain entanglement after several kicks.

In addition to the dynamics, we also check the distribution of the 
momentum of the quantum particle,
\begin{equation}
f_Q(p_1)=\int \int dq_1 dq_2
|\Phi(q_1, q_2)|^2 \delta(p_1-{\partial S \over \partial q_1}).
\end{equation}
By definition, it may differ from the amplitude of 
the wave function in the $p_1$-representation (Fig.2).
The result is given in Fig.7.

%
\section{Summary and Discussion}
%
In order to recognize the role of nonlinearity in quantum systems, 
we estimated quantum trajectories for the coupled kick rotor model.
It has recently been known phenomenologically 
that effect of the nonlinearity appears differently, 
depending on the degrees of freedom of a system
\cite{Adachi88}-\cite{Sakagami96}.
In a single kicked rotor, 
for example, suppression of the momentum diffusion 
appears and the wave function is localized.
For the coupled kick rotors, on the other hand, 
the diffusion in the momentum space seems to continue. 
In this paper, we compared the single rotor with the coupled rotor model 
from the viewpoint of quantum trajectories.
For a single kick rotor, the acceleration of a quantum particle induced 
by the impulsive kicks is suppressed by the quantum potential(Fig. 3(a)).
For the coupled two rotors, by contrast, 
entanglement in the two dimensional configuration space is realized 
(Fig. 6) and suppression of the acceleration by the kick is mild 
in spite of repulsion between trajectories(Fig. 3(c)).
In fact, some particles are really accelerated (Fig. 5).
The dispersion of the momentum of the particles is expanded by the kicks  
and localization is broken for the wave function (Fig. 1).

A bundle of quantum trajectories flow according to 
the continuity equation and each trajectory does never go across the 
others due to the uniqueness theory for solutions of differential equations
\cite{Holland93}.
In one dimensional system, this property appears as the phenomenon 
that the quantum trajectories are repulsive and the complicated 
density distribution tends to be flattened.
Therefore, the relative acceleration between quantum particles
induced by nonlinearity is suppressed in the system.
This results in the localization of the wave function. 
For the system with redundant degrees of freedom like 
our coupled kick rotor model, 
the complicated density distribution is flattened also
into the second degree of freedom.
That is, quantum particles can be scattered forwardly with each other.
As a result, the repulsive nature due to quantum interference 
is not straightforward.
It is the reason why introduction of the coupling yields 
the possibility that some particles can be accelerated and 
the localization of the wave function is broken.

In the de Broglie-Bohm interpretation, 
the momentum of a quantum particle with a rigid trajectory is equalized 
to the spatial gradient of the phase factor in the wave function.
In the pp-coupling model, 
\begin{eqnarray}
\ddot q_2
&=&\dot p_2 +c_{pp}\dot p_1 
=-{\partial \over \partial q_2}V_Q
 -c_{pp}{\partial \over \partial q_1}V_Q \nonumber \\
&=&{\hbar^2 \over 2}{\partial \over \partial q_2}
{1 \over R}({\partial^2 R\over \partial q_1^2}
+{\partial^2 R\over \partial q_2^2}
+2c_{pp}{\partial^2 R\over \partial q_1 \partial q_2})
+{\hbar^2 c_{pp}  \over 2}{\partial \over \partial q_1}
{1 \over R}({\partial^2 R\over \partial q_1^2}
+{\partial^2 R\over \partial q_2^2})
+O(c_{pp}^2). \nonumber \\
& &
\label{dotdotq}
\end{eqnarray}
Equation (\ref{dotdotq}) shows that, 
even if the degrees of freedom, $q_1$ and $q_2$ are uncorrelated initially, 
that is, $R(q_1,q_2)=R(q_1)R(q_2)$, the coupling induces the 
gradient of quantum potential with respect to the $q_1$-direction 
and accelerate particles in the direction of the second degree of freedom, 
$q_2$.
This effect decreases the gradient of the density of quantum particles 
and reduces the repulsion in the $q_1$-direction.
In the work by Adachi et al \cite{Adachi88}, 
the instantaneous coupling between the configuration variables 
$q_1$ and $q_2$ is introduced as a model to analyze.
In this case,
mixing of the phase factor in the wave function is induced directly 
by the coupling and causes the acceleration along the
$q_2$-axis.

Redundant degrees of freedom help the classical diffusion, 
which is analogous to the ergodicity although the phase space is not 
compact in our model.
Our analysis also shows that multi-dimensional systems 
emulate partly the classical dynamics. 
On one hand, the quantum trajectory is altered by repulsion 
in short range.
In global range, on the other hand, it may express nonlinearity 
in the system.
We may say that the Ehrenfest time scale is not active in our analysis 
where the wave function is spread widely from the 
beginning and continues diffusion in the uncompact phase space.
Therefore, there is a possibility that 
we can treat the appearance of the classical property as a matter of 
the scale of the discrepancy between trajectories 
rather than a matter of the time scale, such as the Ehrenfest time scale.
To clarify the dependence of the dynamics on the spatial discrepancy 
between quantum trajectories, 
we have to make further qualitative analysis.
The analysis  need more computational 
memory and more CPU time and will be presented elsewhere.


As demonstrated in this paper, the de Broglie-Bohm picture which 
gives rigid trajectories 
is helpful to understand the dynamics equivalent to the unitary evolution 
driven by the Schr\"{o}dinger equation. 
In the quantum trajectory picture, however, the superposition of 
the right going and left going waves which have the same 
and opposite momentum 
is identified with the static state with zero momentum.
In our analysis, therefore, 
we utilized the averaged velocity $V_i$ rather than 
the velocity at a moment, (\ref{velocity}).
In Fig. 2, the characteristic plateaus appears at the classically 
forbidden point $(q,p)=(\pi,2n\pi)(n=0, \pm 1, \pm 2,...)$ 
which corresponds to the elliptical fixed point, while 
does not in Fig. 7.
For comparison with the statistical distribution on the phase space, 
the Wigner and Husimi descriptions have certainly  
preferable points\cite{Lee93}.
To understand the quantum dynamics in multi-dimensional non-linear systems 
further, 
the de Broglie-Bohm approach and complementally 
the Wigner and Husimi functions will be needed.


%
%
\begin{figure}
\caption{
The time evolution of the dispersion of the momentum, the second order 
moment $Q\equiv<(\hat p_1-<\hat p_1>)^2>$ is shown.
The lines (No) and (P-P) indicate the quantum single kick rotor with 
the kick strength $k_1=2.0$ and the kick rotors with the momentum coupling, 
respectively.
For the coupled kicked rotors, the kick strength $k_1$ and $k_2$ are set 
to 2.0 and 0.9, respectively, 
and the coupling strength $c_{pp}$ 
is 0.2.
The initial state is set to the momentum eigenstate with 
${p_1}_0={p_2}_0=\pi/2$ in Eq. (\ref{Initial}).
}
\label{Fig.1}
\end{figure}
\begin{figure}
\caption{
The momentum distribution $|<p_1|\Phi>|^2$ at $t=100{\rm T}+0$ are shown.
The abbreviations (No) and (P-P) indicate the single kick rotor and 
pp-coupling models, respectively.
The model parameters $k_1$, $k_2$, and $c_{pp}$ are the same as in Fig. 1.
}
\end{figure}
\begin{figure}
\caption{
The quantum trajectory during the time interval from $t={\rm T}$ to 
$t=2{\rm T}$ is shown.
The horizontal axis indicates the time $t$, and 
the vertical one shows the $q_1$-coordinate.
Fig. 3(a) shows the 20 typical trajectories which are set uniformly 
along the $q_1$-coordinate at $t={\rm T}$ for a single rotor.
By contrast, Figs. 3(b) and 3(c) show 
the 20 quantum trajectories which are equally spaced on the surfaces
 $q_2=\pi$ and $q_2=0$, respectively, at $t={\rm T}$.
The model parameters $k_1$, $k_2$, and $c_{pp}$ are the same as in Fig. 1.
}
\end{figure}
\begin{figure}
\caption{
For the coupled kicked rotors, 
the time evolution of probe particles is shown 
in the time interval from $t={\rm T}$ to $t=2{\rm T}$.
At $t={\rm T}$, 1000 quantum particles are ranged homogeneously 
on the surfaces $q_2=\pi$ and $q_2=0$.
The horizontal and vertical axes show the coordinates $q_1$ and $q_2$ 
of the probes, respectively.
Each row like a line shows the snap shot at 
$t=1.0{\rm T},~1.1{\rm T},~1.2{\rm T}, \cdots,~2.0{\rm T}$. 
The model parameters $k_1$, $k_2$, and $c_{pp}$ are the same as 
in Fig. 1.
}
\label{Fig.4}
\end{figure}
\begin{figure}
\caption{
The effective acceleration $F_1(30)$ for $500 \times 20$ probes which 
are settled at $t=30{\rm T}$ is shown.
Figs. 5(a) and 5(b) correspond to the single kick rotor and pp-coupling models, respectively.
For the single kick rotor, the probes are distributed uniformly 
between $0$ and $2\pi$ along the $q_1$-axis.
For the coupled kicked rotors, 
20 rows which consists of 500 probes along the $q_1$ axis 
are arranged uniformly between 0 and $2\pi$.
For each probe, the averaged velocities $V_1(29)$ and $V_2(30) $ are 
estimated by the two computational runs which are different in the 
time step size
($\Delta t=2.5\times10^{-6}{\rm T}$ and $5.0\times10^{-6}{\rm T}$).
For the coupled kicked rotor, 
the 101 probes whose deviations in $V_1(29)$ and $V_2(30)$ are greater than 
$0.1[1/{\rm T}]$ are rejected in Fig. 5(b). 
The model parameters $k_1$, $k_2$, and $c_{pp}$ are the same as in Fig. 1.
}
\label{Fig.5}
\end{figure}
\begin{figure}
\caption{
The positions of $5000\times 10$ probes are shown to visualize 
the deformation of the volume element in the 2-dimensional configuration 
space $(q_1,q_2)$.
Fig. 6(a) shows the initial positions of the probes distributed uniformly 
on the surface $q_2=2 \pi \times i/10 (i=0, 1, \cdots 9)$ at $t={\rm T}$.
Fig. 6(b) and (c) are the snap shots for the positions of the probes
at $t=2{\rm T}$ and $3{\rm T}$, respectively.
The model parameters $k_1$, $k_2$, and $c_{pp}$ are the same as 
in Fig. 1.
}
\label{Fig.6}
\end{figure}
\begin{figure}
\caption{
The distribution of the momentum of the quantum particles are shown.
The distribution function is estimated by 
$f_Q(p_1)=\int \int |\Phi|^2 \delta(p_1-{\partial S \over \partial q_1})
{\rm d}q_1{\rm d}q_2$,
where $\Phi$ is the wave function at $t=100{\rm T}$ for the pp-coupling 
model.
The model parameters $k_1$, $k_2$, and $c_{pp}$ are the same as 
in Fig. 1.
}
\label{Fig.7}
\end{figure}

%
%

\end{document}